\documentclass[aps,preprint]{revtex4-2}

\usepackage{graphics}
\usepackage{subcaption}
\usepackage{amsmath}
\usepackage{mathtools}
\usepackage{color}

\draft 

\begin{document}


\title{FORC Diagram Features of Co Particles due to Reversal by Domain Nucleation}



\author{Leoni Breth}
\affiliation{University of Continuing Education Krems, Department for Integrated Sensor Systems, 2700 Wr. Neustadt, Austria}
\author{Johann Fischbacher}
\affiliation{University of Continuing Education Krems, Department for Integrated Sensor Systems, 2700 Wr. Neustadt, Austria}
\author{Alexander Kovacs}
\affiliation{University of Continuing Education Krems, Department for Integrated Sensor Systems, 2700 Wr. Neustadt, Austria}
\author{Harald Özelt}
\affiliation{University of Continuing Education Krems, Department for Integrated Sensor Systems, 2700 Wr. Neustadt, Austria}
\author{Thomas Schrefl}
\affiliation{University of Continuing Education Krems, Department for Integrated Sensor Systems, 2700 Wr. Neustadt, Austria}
\author{Christoph Czettl}
\affiliation{CERATIZIT Austria GmbH, R \& D Cutting Tools, 6600 Reutte, Austria}
\author{Saskia Kührer}
\affiliation{CERATIZIT Austria GmbH, R \& D Cutting Tools, 6600 Reutte, Austria}
\author{Julia Pachlhofer}
\affiliation{CERATIZIT Austria GmbH, R \& D Cutting Tools, 6600 Reutte, Austria}
\author{Maria Schwarz}
\affiliation{CERATIZIT Austria GmbH, R \& D Cutting Tools, 6600 Reutte, Austria}
\author{Hubert Brückl}
\affiliation{University of Continuing Education Krems, Department for Integrated Sensor Systems, 2700 Wr. Neustadt, Austria}


\date{\today}

\begin{abstract}
First Order Reversal Curve (FORC) diagrams are a popular tool in geophysics and materials science for the characterization of magnetic particles of natural and synthetic origin. However, there is still a lot of controversy about the rigorous interpretation of the origin of certain features in a FORC diagram. In this study, we analyze FORCs computed by micromagnetic simulations of Co cubes with dimensions of 50, 100 and 150~nm and uniaxial magnetocrystalline anisotropy. For the larger cubes we observe the formation of a stable two-domain state. The nucleation of a reversed domain and its subsequent annihilation are clearly visible as separate peaks in the FORC diagram. They spread out along the coordinate axis in the FORC diagram, which is associated with the bias field $H_U$ of a Preisach hysteron. Based on our findings, we state that a FORC diagram peak spreading along the $H_U$ axis can have its origin in the step-wise magnetization reversal driven by nucleation of domains in a single particle. This means that we have identified another mechanism apart from the well-known magnetostatic interaction between a set of particles that leads to features in the FORC diagram extending along the $H_U$-axis. Our study demonstrates that if FORCs shall be used as a quantitative tool to assess the microstructure of samples containing magnetic material, more information from other methods will be required to identify the correct physical mechanism by which a certain ``fingerprint'' in a FORC diagram is produced.
\end{abstract}


\maketitle 


\section{Introduction}\label{sec:intro}
First Order Reversal Curves (FORCs) are a series of magnetization curves, which represent a set of functions $M_i(H, H_{r,i})$. Here, $H$ is an external magnetic field, which for the $i$th FORC starts at a value $H = H_{r,i}$ and is increased towards a value $+H_{\mathrm{sat}}$ where the magnetization saturates. Each FORC is started at values of $ H_{r,i} > -H_{\mathrm{sat}}$, which are called reversal fields. Because they start at fields smaller than the saturation field they correspond to the reversal branches of minor loops.
FORCs are most commonly measured by using Vibrating Sample Magnetometers (VSMs). They are widely used in the geophysics community as a tool to e.g. identify magnetic particles produced by magnetotactic bacteria in sediment samples, which are information carriers of historical climate conditions~\cite{wagner_2021, egli_2013}. FORCs were also in focus for extracting switching field distributions from magnetic recording media~\cite{winklhofer_2006}. When FORCs are plotted in a FORC diagram as a color-coded map of the second mixed derivative $d^2M/dHdH_r$ they reveal subtle characteristics of the magnetization reversal processes which are not made visible by conventional major loop measurements. Hence, FORC diagrams are a powerful tool to identify the presence of certain magnetic ``fingerprints'' in bulk samples. However, rigorous quantitative interpretation of FORC diagrams to this date remains controversial. \\
Our contribution to this discussion is a detailed study of FORC diagram features arising from well known magnetization states in Co cubes with edge lengths of 50, 100, and 150~nm. We use micromagnetic simulations of FORCs to study the magnetization reversal process and we observe nucleation of reversed domains for cubes of 100 and 150~nm edge lengths. Co inclusions with these dimensions occur in the Co binder matrix of industrially produced cemented carbides such as tungsten carbide (WC). Magnetic characterization by major loop parameters such as the coercive field and the saturation magnetization has a long tradition in this field as a means for quality control~\cite{garcia_2019}. The WC grain size is tuned by the production process, which is crucial to get a cemented carbide with highly customized mechanical properties. Apart from the WC skeleton the morphology of the Co binder phase was also identified to play an important role, but at present can only be analyzed with high-resolution electron microscopy~\cite{eizadjou_2020}. FORCs are expected to provide more insight into the details of the Co binder structure than the current evaluation using major loops. \\
 The paper is organized as follows: First, we revise the present understanding of the FORC method as a mere way to image a ``magnetic fingerprint'' of a sample, but also the attempts to give a formal understanding of the FORC distribution as a distribution of switching entities called hysterons (Section ~\ref{sec:forc}). 
 Details on the setup of the micromagnetic simulations will be explained in Section~\ref{sec:mm}. In Section~\ref{sec:part} we will present the size-dependent features evolving in FORC diagrams, which are driven by domain nucleation and wall propagation. Section~\ref{sec:disc} will discuss our results in context with the results of others and we finish with a conclusion in Section~\ref{sec:conc}.

\section{First Order Reversal Curves (FORCs) and their visualisation}\label{sec:forc}
\begin{figure}
    \centering
    \includegraphics{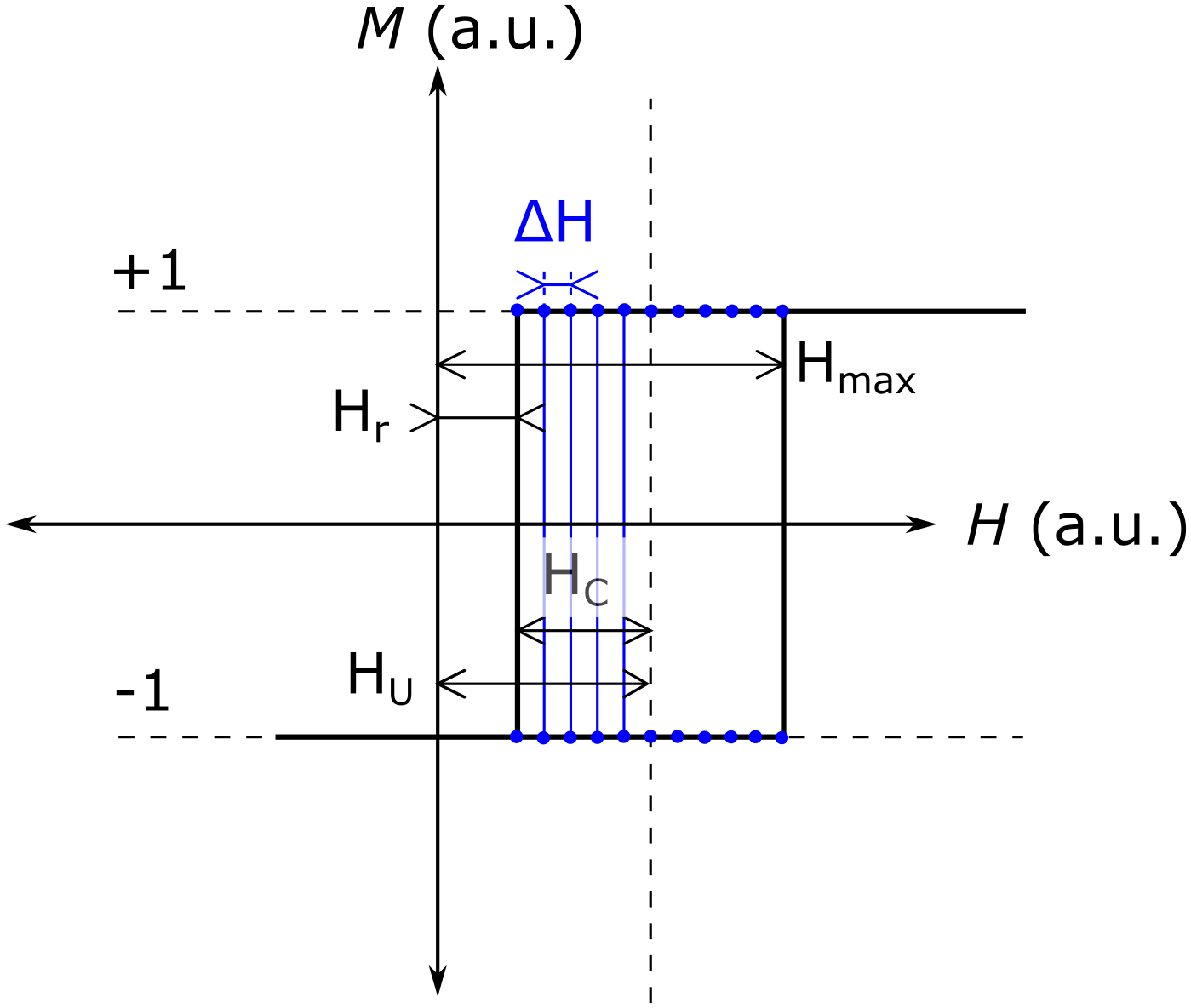}
    \caption{Preisach hysteron with mapping of variables $H_C = \frac{1}{2}(H - H_r)$ and $H_U = \frac{1}{2}(H + H_r)$. On one FORC $H_r$ is fixed and $H$ is increased by steps of $\Delta H$ up to $H_{\mathrm{max}}$. }
    \label{fig:hysteron}
\end{figure}

 It was shown by Mayergoyz~\cite{mayergoyz_1986} that by assuming simple square hysteresis operators (Preisach hysterons~\cite{preisach_1935}) defined by a switching field $H_C$ and a bias field $H_U$ (see Fig.~\ref{fig:hysteron}) a fundamental mathematical model for any kind of hysteresis phenomenon can be derived. The Preisach coordinates $H_C$ and $H_U$ are mapped to the variables from a FORC measurement $H$ and $H_r$ by the following relations:
 \begin{eqnarray}
     H_C & = & \frac{1}{2}(H - H_r), \\
     H_U & = & \frac{1}{2}(H + H_r)
 \end{eqnarray}
 
 When the hysteresis operators $\hat{\gamma}_{\mathrm{H, H_r}}$ are applied to some general time-dependent input signal $u(t)$, this yields
\begin{equation}\label{eq:gamma}
   \hat{\gamma}_{\mathrm{H, H_r}} u(t) = \pm 1,  
\end{equation}

whereby the edges of the transitions from/to $\pm 1$ are located at values of the input signal $u = H_r$ and $u = H$. 
The irreversible part of a magnetization curve (or any other function $f$ describing a process with hysteresis) is then a weighted sum of all Preisach hysterons with widths in the range $[H_r +\Delta H, H_{\mathrm{max}}]$, which in the continuous limit can be written as a double integral

\begin{equation}\label{eq:integral}
    M(H(t)) \equiv f(t) = \int\int_{|H_r| \geq H} \mu(H, H_r) \hat{\gamma}_{\mathrm{H, H_r}} u(t)dHdH_r.
\end{equation}

Further, it turns out using a geometric solution~\cite{mayergoyz_1986} for \eqref{eq:integral} that

\begin{equation}\label{eq:forcdef}
    \mu(H, H_r) \propto \rho(H, H_r) \coloneqq -\frac{1}{2}\frac{\partial^2 M(H, H_r)}{\partial H \partial H_r }
\end{equation}

\eqref{eq:forcdef} is called the FORC distribution, which is proportional to the weight function $\mu(H, H_r)$ for the Preisach hysterons, if the following two conditions hold:

\begin{enumerate}
    \item Minor loops between the same boundary values, but with different magnetic history, always have the same shape and size (\textbf{congruency property}).
    \item When reaching a global energy minimum all memory about previous minima on that reversal path is deleted (\textbf{wiping-out property})\cite{dellatorre_2005}.
\end{enumerate}

The second property corresponds to the well-known phenomenon of ``erasing magnetic history'' by applying a high field beyond saturation. For a wide range of magnetic materials, the congruency property is shown not to be fulfilled, which has triggered extensions to the Classical Preisach Model with increasing complexity~\cite{dellatorre_1966, kadar_1987, mayergoyz_1988}.

Another important physical aspect of the interpretation of Eq.~\ref{eq:integral} is that $\mu(H, H_r) \hat{\gamma}_{\mathrm{H, H_r}} u(t)$ corresponds to a mean state of all hysterons~\cite{egli_2021}. This means that when we obtain the FORC distributions from measurements on bulk samples it will represent an average over all irreversible magnetization processes that occur in the sample. 


However, neither the congruency nor the wiping-out properties are necessary conditions to experimentally determine the FORC distribution given by Eq.~\ref{eq:forcdef}~\cite{pike_1999}. Pike et al.~\cite{pike_2005} successfully fitted FORC distributions of bit patterened magnetic recording media by combining a moving Preisach model with curvilinear hysterons. In their work, they use a phenomenological function to model the hysteresis loop of the individual Ni nanopillars in the recording medium. Obviously, this approach can only work for well-defined systems for which the microscopic magnetic switching entities can be investigated independently by experimental methods such as magnetic force microscopy~\cite{ross_2002}. Whenever such information is not available micromagnetic simulations are a powerful tool to support inferences about the origin of the magnetization changes visible in a FORC diagram. 





\section{Micromagnetic Simulations of FORCs}\label{sec:mm}
\begin{figure}
    \centering
    \includegraphics{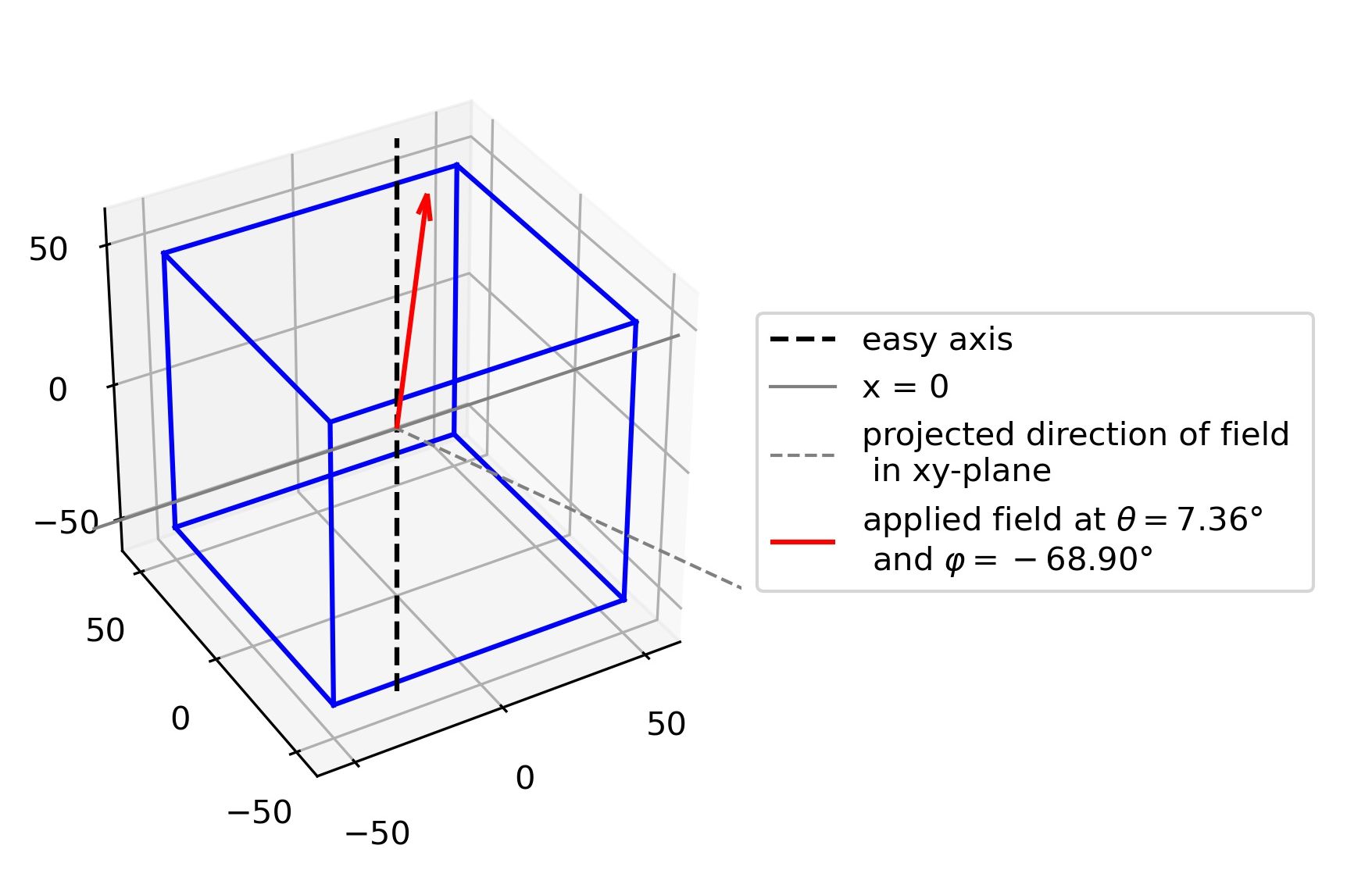}
    \caption{Relative alignment of the applied external field to the particle's magnetic easy axis as well as to the particle's geometry. $\theta = 7.36^{\circ}$ denotes the polar angle and $\varphi = -68.90^{\circ}$ the azimuth angle of the applied field.}
    \label{fig:cube}
\end{figure}

In the geophysics community, micromagnetic simulations are becoming increasingly popular as a means to support inferences with the aid of FORC diagrams on e.g. historical climate conditions. The interpretation of FORC diagrams measured on drill cores containing needle-shaped magnetofossils produced by magnetotactic bacteria~\cite{wagner_2021} was supported by modeling the magnetization reversal of such needle-shaped particles. Also ferrimagnetic framboidal greigite (Fe$_3$S$_4$) was modeled using micromagnetic simulations, which made it possible to show the effect of magnetostatic interaction of closely packed greigite grains on the FORC diagram~\cite{valdez_2020}. Furthermore, it has been shown by micromagnetic simulations that the range of particle sizes hosting magnetization states which are capable of reliably recording the earth's magnetic field extends far beyond the single domain limit, where stable single vortex states are formed~\cite{nagy_2019}.
We use our in-house developed micromagnetic code to model FORCs of hexagonally close packed cobalt (hcp Co) cubes with edge lengths of 50, 100 and 150~nm.  The simulation of FORCs of such large microstructures is made feasible by usage of a nonlinear conjugate gradient method to compute the energy minima of Gibbs' energy~\cite{fischbacher_2017}. We use a quadratic form for Gibbs' energy, which is suitable to simulate hcp Co as a material with uniaxial anisotropy~\cite{schrefl_2007}. Further speed-up of the energy path calculation is achieved by preconditioning using a local approximation of the Hessian representing the minimization problem~\cite{exl_2018}. Co particles with low shape anisotropy, uniaxial magnetic anisotropy and dimensions above the single domain limit to this date have not been studied using FORC diagrams.  For all the simulations presented here, the external field $H$ is applied at a polar angle $\theta = 7.36^{\circ}$ and at an azimuth angle $\varphi = -68.90^{\circ}$ with respect to the easy axis resulting from the uniaxial magnetocrystalline anisotropy for hcp Co. The spatial orientation of the field axis with respect to the cube's as well as the anisotropy axes is shown in Fig.~\ref{fig:cube}. The tuple $\theta$ and $\varphi$ is the first tuple in a series generated by a Halton sequence, that yield uniformly distributed points on the surface of a sphere. The features in the FORC diagram of this particular setup have a general character, as their origin can be fully explained by domain nucleation and annihilation along different paths of the micromagnetic energy landscape.
For the simulation of FORCs, the magnetization states at the starting fields $H_r$ first have to be determined. This is done by calculating the major loop by decreasing the external field $H$ in steps of width $\Delta H$ from positive to negative saturation. Whenever the calculated magnetization has changed by a minimum amount $\Delta M$ with respect to the previous state the configuration of magnetization vectors on the nodes of the finite element mesh is written out. These configurations are then used as a starting configuration for the FORCs. The maximum mesh size was set to 3~nm in order to be able to resolve magnetization changes driven by the competition between magnetostatic and exchange energy ($l_{\mathrm{ex,Co}} = \sqrt{\mu_0 A / J_S^2 } = 3.45$~nm~\cite{Exl_2020}, material parameters are given in Tab.~\ref{tab:simparam}).

\begin{table}[h]
    \centering
    \begin{tabular}{cccccc}
         \hline
         $\mu_0 \Delta H\quad$ & $\mu_0 \Delta M\quad$ & $\mu_0 H_{\mathrm{max}}\quad$ & $\mu_0 M_S\quad$ & $K\quad$ & $A$\\ \hline \hline
         0.02~T $\quad$ & 0.01~T $\quad$ & 2.0~T $\quad$ & 1.81~T $\quad$ & $4.1\cdot 10^5  \mathrm{J/m^3}\quad$  & $3.1\cdot 10^{-11} \mathrm{J/m}$ \\ \hline
    \end{tabular}
    \caption{Summary of simulation and material parameters for hcp Co: $\mu_0 \Delta H$ ... step of external field, $\mu_0 \Delta M$ ... min. change of magnetization when a new state is written out as starting point for FORC,  $\mu_0 H_{\mathrm{max}}$ ... max. external field to which the FORCs return, $\mu_0 M_S$ ... saturation magnetization, $K$ ... uniaxial magnetocrystalline anisotropy constant, $A$ ... micromagnetic exchange stiffness }
    \label{tab:simparam}
\end{table}


\section{Particle-size dependent evolution of FORC diagrams}\label{sec:part}
\begin{figure*}
    \includegraphics[width=0.9\textwidth]{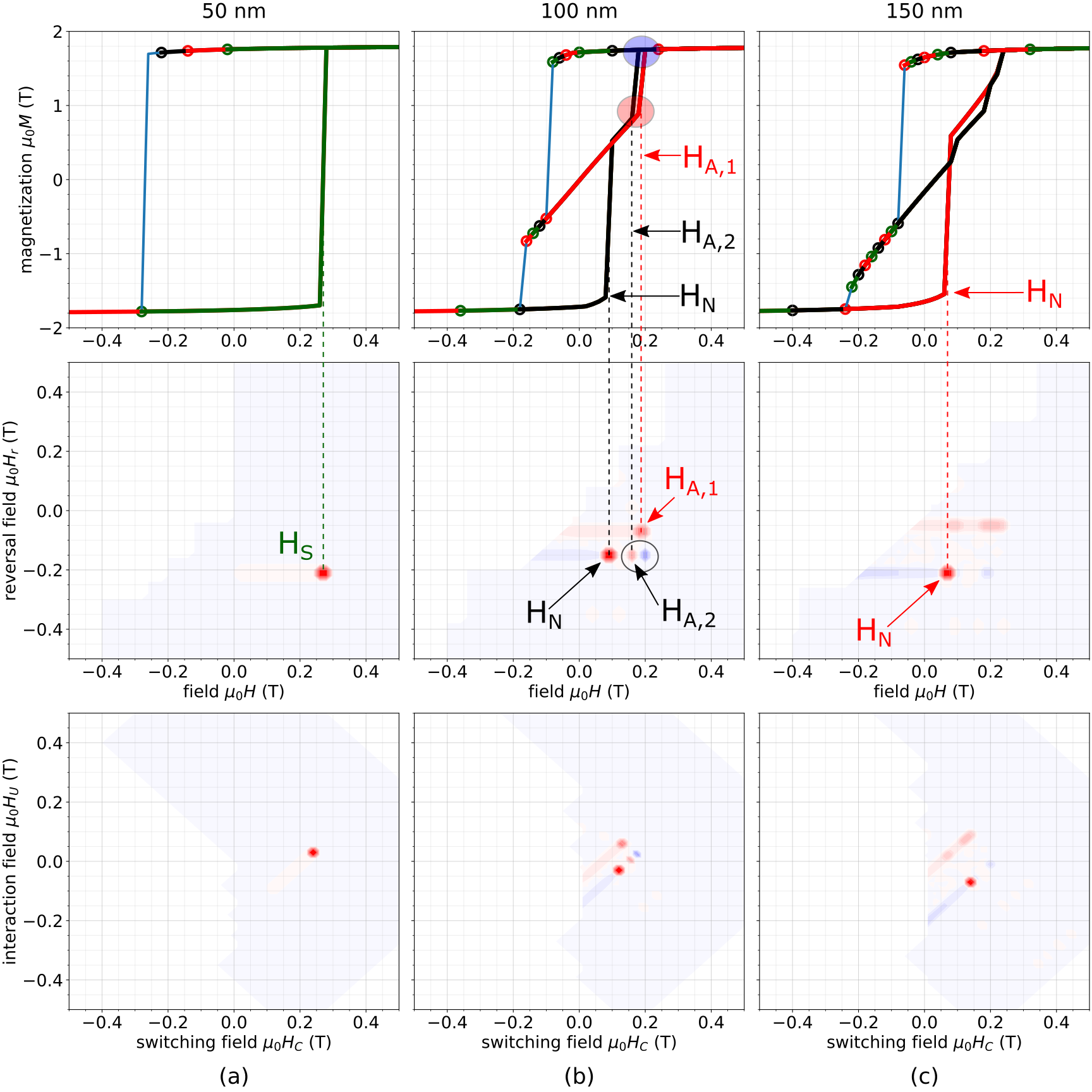}
    \caption{FORCs and FORC diagrams in the regular (reversal field $H_r$ vs applied field $H$) and the Preisach coordinate systems (interaction field $H_U$ vs switching field $H_C$) for hcp Co cubes with varying edge length. (a) 50 nm: The magnetization reverses by coherent rotation (macrospin) with a single defined switching field $H_S$, which is visible as a single peak in the FORC diagrams. (b) 100 nm: A reversed domain is nucleated at a field $H_N = \pm 0.1$~T. Depending on the reversal path (red or black FORCs) the annihilation fields of the domain wall differ $H_{A,1} \neq H_{A,2}$  and two (red i.e. positive) peaks are visible in the FORC diagram. Because the black FORC crosses the red FORC, there is a positive-negative double peak. (c) 150 nm: The nucleation field $H_N$ is still visible as a pronounced peak. The annihilation peak above is smeared out.}  
    \label{fig:ani119}
\end{figure*}

In this section we discuss in detail how certain reversal mechanisms observed in simulations of hcp Co cubes become visible in a FORC diagram. To analyze our magnetization data from micromagnetic simulations we used our own implementation of the central difference scheme in Python to compute a FORC distribution (see Sec.~\ref{sec:comp}).The simulated FORCs as well as the major loops for Co cubes with edge lengths of 50, 100 and 150~nm are shown in the upper row of Fig.~\ref{fig:ani119}. 
The major loop in Fig.~\ref{fig:ani119}a shows a single step at the coercive field of the particle. There are no stable multidomain states along the demagnetization curve. The single peak shown in the FORC diagram below originates from the mixed second derivative calculated according to Eq.~\ref{eq:forccomp} on the FORC starting at $\mu_0 H_r = -0.22$~T. Such a ``Delta function peak'' has been described analytically by Pike and Fernandez for an ideal square hysteresis loop, which can be modeled by a step function~\cite{pike_fernandez_1999}. The step is located at the switching field $H = H_S$ for the 50~nm particle.

\begin{figure*}
    \includegraphics{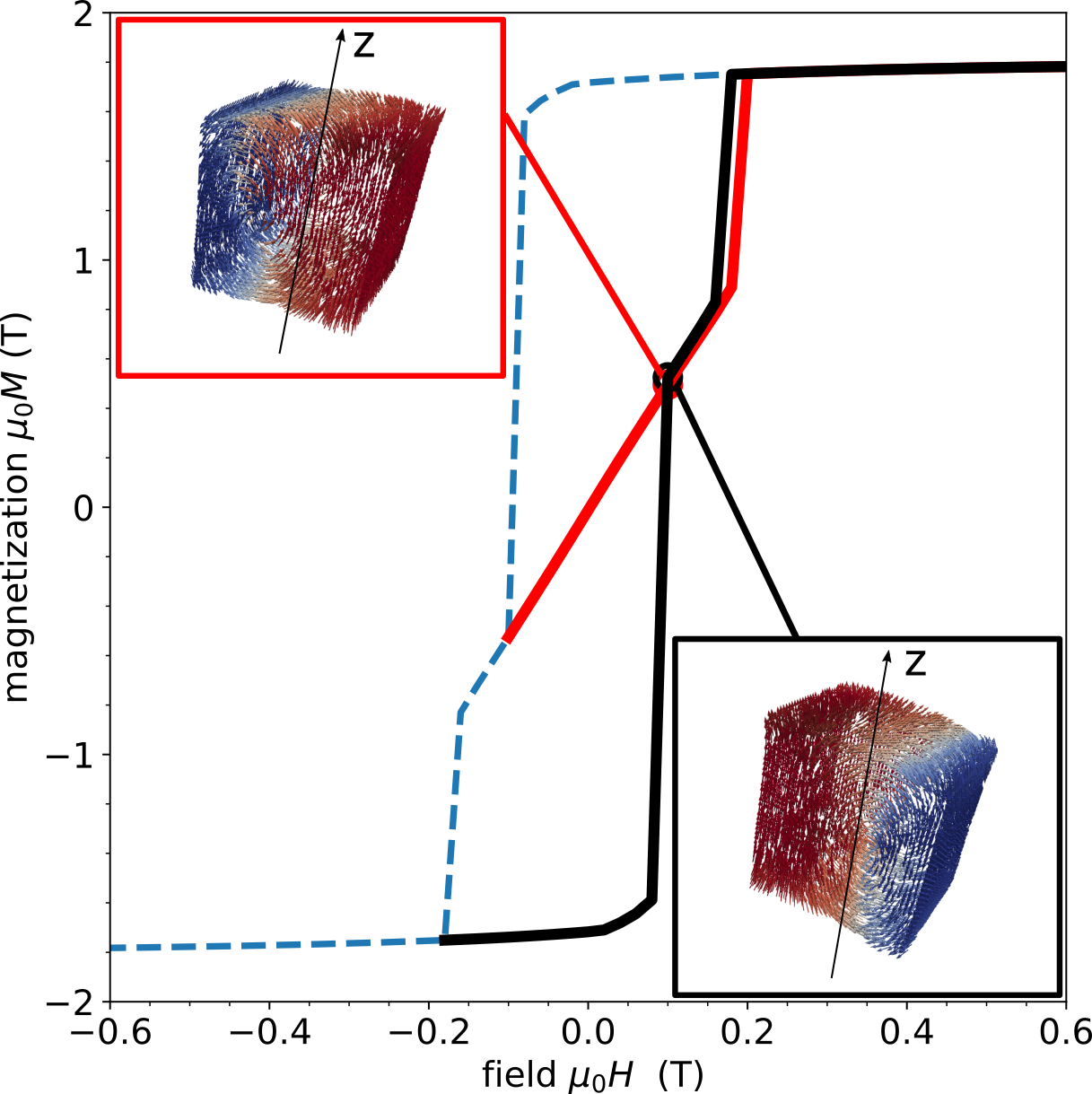}
    \caption{Domain wall nucleation and annihilation on two different FORCs: The annhilation field on the red FORC $H_{A,1}$ is larger than the annihilation $H_{A,2}$ on the black FORC. The vortex visible in the insets is only present at the surface of the cube due to the contribution from demagnetizing field, which favorizes alignment of magnetic moments perpendicular to the easy axis (z axis)~\cite{rave_1998}.}
    \label{fig:100nm}
\end{figure*}

With increasing particle size, the reversal mode changes (Fig.~\ref{fig:ani119}b and c). A reversed domain forms at the domain wall nucleation field $H_N$, which then expands. In contrast to smaller 50~nm Co cube, a multidomain state remains stable for a range of external fields. Only after the external field reaches the critical value at the annihilation field $H_A$ (see Fig.~\ref{fig:100nm}), the domain wall is moved out of the sample and the magnetization is fully reversed.

For the Co cubes with 100 and 150~nm edge lengths, the FORCs starting in this two-domain state show a higher annihilation field $H_{A,1}$ than the FORCs starting from negative saturation, i.e. $H_{A,1} > H_{A,2}$. Detailed investigation of the reversal process reveals that the difference in the annihilation fields results from different reversal paths dependent on the initial state of the FORC. The domains which are nucleated on the FORC starting at negative saturation show opposite symmetry to the initial state of the FORC starting in the two-domain state. The two-domain states are shown in Fig.~\ref{fig:100nm} just before annihilation. Because these two FORCs cross each other there is a double positive-negative peak in the FORC diagram additionally to the two positive peaks and their respective positive and negative ``tails''. 

\section{Discussion}\label{sec:disc}
A FORC diagram similar to the one for the 100~nm cube reported here was observed experimentally for an array of permalloy stripes patterned out of a 50~nm thin film with widths ranging from 10 to 100~$\mathrm{\mu m}$~\cite{gross_2019}. The occurrence of a double peak in the FORC diagrams presented there was attributed to magnetostatic interaction and hence called ``interaction peak''.

\begin{figure*}
    \includegraphics[width=0.9\textwidth]{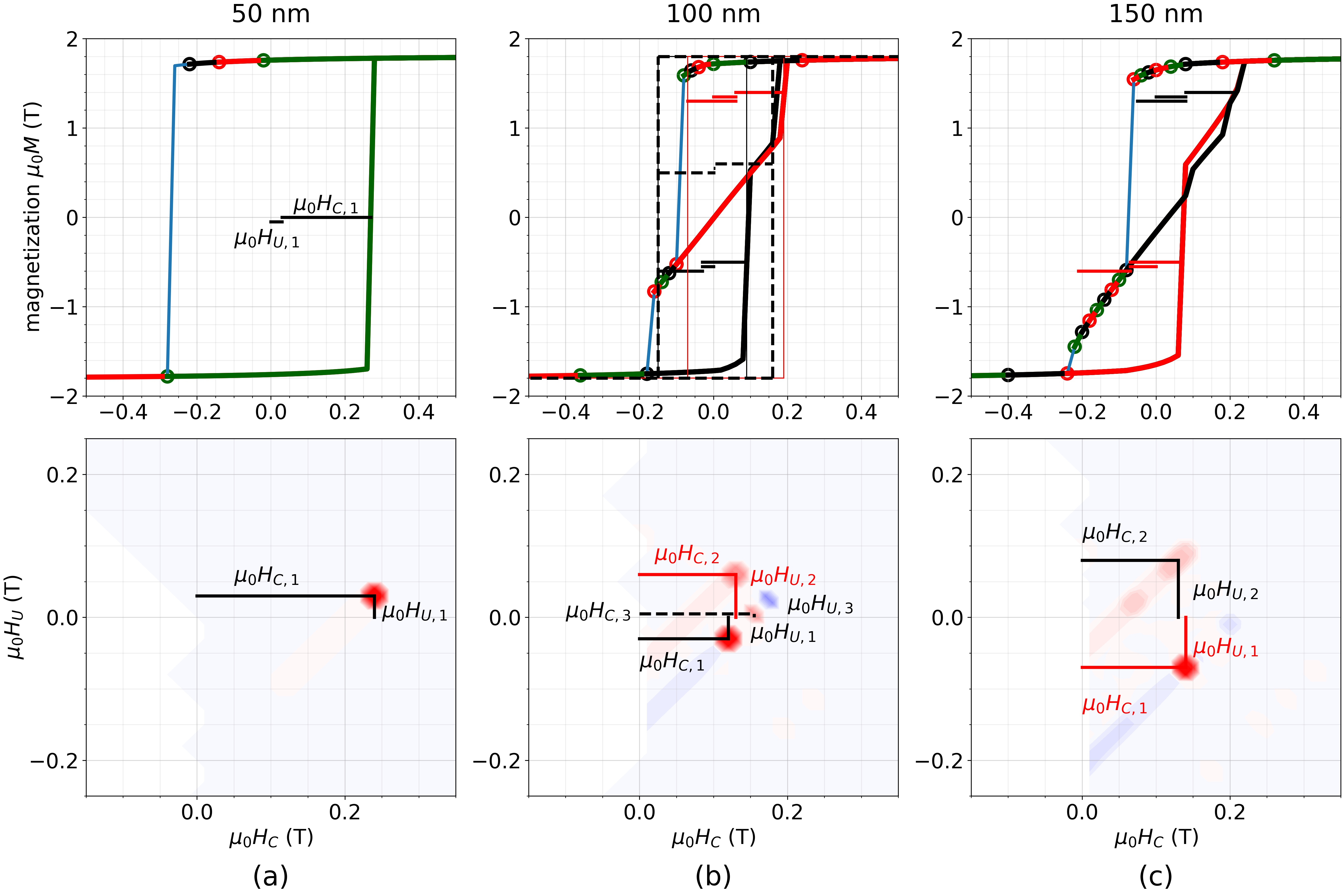}
    \caption{FORCs and FORC diagrams in the regular and the Preisach coordinate systems for cube edge lengths of (a) 50 nm, (b) 100 nm and (c) 150 nm. The peak coordinates ($H_{C,i}$,$H_{U,i}$) (see also Tab.~\ref{tab:HcHu}) are the defining parameters of the hysterons representing the irreversible magnetization jumps.}  
    \label{fig:ani119_HcHu}
\end{figure*}

Our results show, however, that such double peaks can also occur in FORC diagrams of individual particles hosting stable multidomain states, without the presence of interparticle interactions. 
For each point feature with a non-negative value in the FORC diagram we can inscribe a hysteron into the simulated FORCs (see Fig.~\ref{fig:ani119_HcHu}, $H_{C,i}$ and $H_{U,i}$ are represented by the horizontal bars in the top row.). The hysteron for the 100~nm cube with a switching field $\mu_0 H_{C,3}$ and a bias field $\mu_0 H_{U,3}$, which corresponds to the positive part of the double peak, appears additionally to the two more prominent peaks from the upper and lower part of the hysteresis loop. For the hysterons derived from the data in Fig.~\ref{fig:ani119_HcHu}~b), we can see that the following relations apply with respect to the nucleation and annihilation fields $H_N$, $H_{A,1}$ and $H_{A,2}$:
\begin{eqnarray}
    H_N  = H_{C,1} - H_{U,1} \\
    H_{A,1} = H_{C,2} + H_{U,2} \\
    H_{A,2} = H_{C,3} + H_{U,3}.
\end{eqnarray}

These results resemble those derived analytically before for the vortex in a Co nanopillar with two different annihilation fields~\cite{pike_1999}. As for the vortex, also for the domain wall a non-negative feature close to the $H_U = 0$-axis pops up in the FORC diagram, which is of particular interest if FORC diagrams are used to detect single-domain magnetic particles. \\

\begin{table}[h]
    \centering
    \begin{tabular}{c||c|c|c}
         & 50~nm & 100~nm & 150~nm \\ \hline \hline
         $\mu_0 H_{C,1}$ & 0.240 & 0.120 & 0.140 \\ \hline
         $\mu_0 H_{U,1}$ & 0.030 & -0.030 & -0.070 \\ \hline
         $\mu_0 H_{C,2}$ & - & 0.130 & 0.130 \\ \hline
         $\mu_0 H_{U,2}$ & - & 0.060 & 0.080 \\ \hline
         $\mu_0 H_{C,3}$ & - & 0.155 & - \\ \hline
         $\mu_0 H_{U,3}$ & - & 0.005 & - 
    \end{tabular}
    \caption{Overview of the Preisach hysteron coordinates in Fig.~\ref{fig:ani119_HcHu}.}
    \label{tab:HcHu}
\end{table}

Finally, we would like to stress that calling $H_U$ an ``interaction field'' does not cover all the physics that lead to features in a FORC diagram extending along the $H_U$ coordinate. The understanding of $H_U$ as a measure of magnetostatic interaction between particles certainly has its validity for samples, where additional evidence is available that the magnetic constituents are within a certain size range, which stays below the single domain limit~\cite{pike_1999, muxworthy_2004, stancu_2003, pike_2005}. However, it has also been shown by analytical models that random domain wall pinning processes can produce features in a FORC diagram extending along the $H_U$-axis, which were found suitable to explain FORC diagrams of annealed transformer steels, but not of geological samples containing multidomain particles~\cite{pike_2001}. Temperature-dependent measurements on multidomain magnetite powders showed no dependence of the peak distribution along $H_U$, as would be expected if the peak width was predominantly due to magnetostatic interactions, which scale inversely to $M_S(T)$~\cite{dunlop_1990}. Together with our study, it becomes clear, that a rigorous quantitative interpretation of FORC diagrams requires further efforts in understanding the underlying physics of the possible magnetization reversal and particle interaction processes that lead to the emergence of specific features~\cite{lascu_2018, roberts_2017, carvallo_2003}.

\section{Conclusion}\label{sec:conc}
In this paper we presented FORC diagrams obtained from micromagnetic simulations of Co cubes with uniaxial anisotropy and dimensions extending beyond the single domain limit. The FORC diagrams show point features spreading along the $H_U$ coordinate, which corresponds to the bias field of Preisach hysterons. This bias is commonly assumed to have its physical origin in interparticle interactions, which is well documented for samples consisting of single domain particles only. It is less clear, however, how this spreading along $H_U$ arises in FORC diagrams of samples dominated by particles with stable multidomain states. By simulating Co cubes with dimensions above the single domain limit we show that peaks arising from magnetization jumps due to nucleation of domains and subsequent annihilation spread out along the $H_U$ axis. Unlike the common interpretation we state that in our case this spreading cannot be due to interparticle magnetostatic interactions, because the simulation is based exclusively on one single Co cube. Based on the increasing spread along $H_U$ with increasing particle size we do acknowledge though that the demagnetization field and hence some ``intra-particle'' magnetostatic interaction might be relevant. This will be the subject of future investigations. Our results show, that FORCs and FORC diagrams are inherently ambiguous when we want to use them for inferences about the sample's microstructure. However, micromagnetic simulations based on local microstructural data can serve as a powerful tool to narrow down the solution space and will make it possible to identify the most likely sources producing a magnetic signal.


%
%

%

\begin{acknowledgments}
The financial support by the Austrian Federal Ministry of Climate Action, Environment, Energy, Mobility, Innovation and Technology (BMK) in the KI-Carbide project \#877141 is gratefully acknowledged.
\end{acknowledgments}

\begin{appendix}
\section{Computing FORC Distributions}\label{sec:comp}

The FORC distribution (see also Eq.~\ref{eq:forcdef})
\begin{equation}\label{eq:forcdist}
    \rho(H, H_r) = -\frac{1}{2}\frac{\partial^2 M(H, H_r)}{\partial H \partial H_r }
\end{equation}

is defined as a second mixed derivative of the surface $M(H, H_r)$ spanned by all FORCs. This means, that regardless of the nature of the FORCs --- be they experimentally measured or simulated --- a higher derivative has to be computed numerically, as each $i$-th FORC is available as a series of discrete data points $M_i(H, H_{r,i})$ forming a surface in 3d space, whereby $H \in [H_r, H_{\mathrm{max}}]$. Pike et al.~\cite{pike_1999} suggested to perform the calculation of $\rho(H, H_r)$ and smoothing of the raw data by performing a local polynomial approximation on every $(2S+1)(2S+1)$ data points , i.e.
\begin{equation}
    M(H, H_r) \approx a_1 + a_2H + a_3 H^2 + a_4H_r + a_5H_r^2 + a_6HH_r \quad \Rightarrow \quad \rho_{\mathrm{local}}(H, H_r) \approx a_6.
\end{equation}

$S$ defines the number of data points in each direction of the $H$ and $H_r$ coordinates and is called smoothing factor, because the more data points are used for the local approximation the more short-range variations of $M$ will be smoothed out. This is useful for noisy data, but there is a trade-off between noise removal and information loss. The widely used software package FORCinel~\cite{harrison_forcinel_2008} provides a method to optimize the choice of the smoothing factor. Another popular software in the FORC community is VARIFORC, which performs the polynomial expansion on the rotated Preisach plane coordinates $H_C$ and $H_U$ and introduces anisotropic smoothing in perpendicular directions~\cite{egli_variforc_2013}. More recently, xFORC~\cite{zhao_2017} (requires LabView 2012), FORC+ (Windows only standalone program)~\cite{visscher_2019}, Forcot (open source MATLAB code with support for Windows and Mac OS X platforms)~\cite{berndt_2019} and doFORC (standalone application on Windows, requires Wine in Unix environments)~\cite{cimpoesu_2019} were released for public use. We used our own code written in Python 3.6 using only publicly available libraries to compute the FORC distribution $\rho(H, H_r)$ by finite central differences without any additional smoothing. The second mixed derivative is computed pointwise using the $k \times l$ array of datapoints for the magnetization $M_{k,l}$, whereby the row index $k$ refers to the reversal field coordinate $H_r$ and the column index $l$ refers to the coordinate for the applied field $H$. The coordinates $H$ and $H_r$ are assumed to be spaced regularly by steps $\Delta H$ and $\Delta H_r$, respectively. Then, applying the central difference scheme $f'(x) \approx [f(x + h) - f(x - h)]/2h$ to both coordinates yields

\begin{equation}\label{eq:forccomp}
    \left( \frac{\partial^2 M}{\partial H \partial H_r} \right)_{k,l} \approx \frac{1}{4\Delta H \Delta H_r}
    \left( M_{k-1, l-1} - M_{k+1, l-1} - M_{k-1, l+1} + M_{k+1, l+1}\right).
\end{equation}

Eq.~\ref{eq:forccomp} involves only summation and subtraction of floating point numbers instead of the necessity to fit a second order polynomial in each step. Only minimal smoothing is performed, which is inherent to the central difference scheme. The same approach is also used in FORC+~\cite{visscher_2019}. 
The most important difference between simulation and experimental data is the fact that the simulation data is sparse along the $H_r$ coordinate. This means, that there is not a new FORC starting at every value of the external field $H$. Only if there is a significant change of the magnetization $\Delta M$, a new FORC is started automatically at this value of $H = H_r$. The values of $H_r$ are found by first simulating the major loop. The simulated FORC data are then transferred to a regular grid of $201 \times 201$ points of $H$ and $H_r$, respectively. The values of $H_r$ where no FORC is simulated are padded with the values of the previous FORC. This ensures, that the algorithm defined by Eq.~\ref{eq:forccomp} can make use of a constant $\Delta H_r$.



    

\end{appendix}

\bibliography{refs}

\providecommand{\noopsort}[1]{}\providecommand{\singleletter}[1]{#1}%
\begin{thebibliography}{37}%
\makeatletter
\providecommand \@ifxundefined [1]{%
 \@ifx{#1\undefined}
}%
\providecommand \@ifnum [1]{%
 \ifnum #1\expandafter \@firstoftwo
 \else \expandafter \@secondoftwo
 \fi
}%
\providecommand \@ifx [1]{%
 \ifx #1\expandafter \@firstoftwo
 \else \expandafter \@secondoftwo
 \fi
}%
\providecommand \natexlab [1]{#1}%
\providecommand \enquote  [1]{``#1''}%
\providecommand \bibnamefont  [1]{#1}%
\providecommand \bibfnamefont [1]{#1}%
\providecommand \citenamefont [1]{#1}%
\providecommand \href@noop [0]{\@secondoftwo}%
\providecommand \href [0]{\begingroup \@sanitize@url \@href}%
\providecommand \@href[1]{\@@startlink{#1}\@@href}%
\providecommand \@@href[1]{\endgroup#1\@@endlink}%
\providecommand \@sanitize@url [0]{\catcode `\\12\catcode `\$12\catcode
  `\&12\catcode `\#12\catcode `\^12\catcode `\_12\catcode `\%12\relax}%
\providecommand \@@startlink[1]{}%
\providecommand \@@endlink[0]{}%
\providecommand \url  [0]{\begingroup\@sanitize@url \@url }%
\providecommand \@url [1]{\endgroup\@href {#1}{\urlprefix }}%
\providecommand \urlprefix  [0]{URL }%
\providecommand \Eprint [0]{\href }%
\providecommand \doibase [0]{https://doi.org/}%
\providecommand \selectlanguage [0]{\@gobble}%
\providecommand \bibinfo  [0]{\@secondoftwo}%
\providecommand \bibfield  [0]{\@secondoftwo}%
\providecommand \translation [1]{[#1]}%
\providecommand \BibitemOpen [0]{}%
\providecommand \bibitemStop [0]{}%
\providecommand \bibitemNoStop [0]{.\EOS\space}%
\providecommand \EOS [0]{\spacefactor3000\relax}%
\providecommand \BibitemShut  [1]{\csname bibitem#1\endcsname}%
\let\auto@bib@innerbib\@empty
\bibitem [{\citenamefont {Wagner}\ \emph {et~al.}(2021)\citenamefont {Wagner},
  \citenamefont {Egli}, \citenamefont {Lascu}, \citenamefont {Lippert},
  \citenamefont {Livi},\ and\ \citenamefont {Sears}}]{wagner_2021}%
  \BibitemOpen
  \bibfield  {author} {\bibinfo {author} {\bibfnamefont {C.~L.}\ \bibnamefont
  {Wagner}}, \bibinfo {author} {\bibfnamefont {R.}~\bibnamefont {Egli}},
  \bibinfo {author} {\bibfnamefont {I.}~\bibnamefont {Lascu}}, \bibinfo
  {author} {\bibfnamefont {P.~C.}\ \bibnamefont {Lippert}}, \bibinfo {author}
  {\bibfnamefont {K.~J.~T.}\ \bibnamefont {Livi}},\ and\ \bibinfo {author}
  {\bibfnamefont {H.~B.}\ \bibnamefont {Sears}},\ }\bibfield  {title} {\bibinfo
  {title} {In situ magnetic identification of giant, needle-shaped
  magnetofossils in {Paleocene}–{Eocene} {Thermal} {Maximum} sediments},\
  }\bibfield  {journal} {\bibinfo  {journal} {PNAS}\ }\textbf {\bibinfo
  {volume} {118}},\ \href {https://doi.org/10.1073/pnas.2018169118}
  {10.1073/pnas.2018169118} (\bibinfo {year} {2021}),\ \bibinfo {note}
  {publisher: National Academy of Sciences Section: Physical
  Sciences}\BibitemShut {NoStop}%
\bibitem [{\citenamefont {Egli}\ \emph {et~al.}(2013)\citenamefont {Egli},
  \citenamefont {Florindo},\ and\ \citenamefont {Roberts}}]{egli_2013}%
  \BibitemOpen
  \bibfield  {author} {\bibinfo {author} {\bibfnamefont {R.}~\bibnamefont
  {Egli}}, \bibinfo {author} {\bibfnamefont {F.}~\bibnamefont {Florindo}},\
  and\ \bibinfo {author} {\bibfnamefont {A.~P.}\ \bibnamefont {Roberts}},\
  }\bibfield  {title} {\bibinfo {title} {Introduction to 'magnetic iron
  minerals in sediments and their relation to geologic processes, climate, and
  the geomagnetic field'},\ }\href
  {https://doi.org/https://doi.org/10.1016/j.gloplacha.2013.10.009} {\bibfield
  {journal} {\bibinfo  {journal} {Global and Planetary Change}\ }\textbf
  {\bibinfo {volume} {110}},\ \bibinfo {pages} {259} (\bibinfo {year}
  {2013})},\ \bibinfo {note} {magnetic iron minerals in sediments and their
  relation to geologic processes, climate, and the geomagnetic
  field}\BibitemShut {NoStop}%
\bibitem [{\citenamefont {Winklhofer}\ and\ \citenamefont
  {Zimanyi}(2006)}]{winklhofer_2006}%
  \BibitemOpen
  \bibfield  {author} {\bibinfo {author} {\bibfnamefont {M.}~\bibnamefont
  {Winklhofer}}\ and\ \bibinfo {author} {\bibfnamefont {G.~T.}\ \bibnamefont
  {Zimanyi}},\ }\bibfield  {title} {\bibinfo {title} {Extracting the intrinsic
  switching field distribution in perpendicular media: {A} comparative
  analysis},\ }\href {https://doi.org/10.1063/1.2176598} {\bibfield  {journal}
  {\bibinfo  {journal} {Journal of Applied Physics}\ }\textbf {\bibinfo
  {volume} {99}},\ \bibinfo {pages} {08E710} (\bibinfo {year} {2006})},\
  \bibinfo {note} {publisher: American Institute of Physics}\BibitemShut
  {NoStop}%
\bibitem [{\citenamefont {García}\ \emph {et~al.}(2019)\citenamefont
  {García}, \citenamefont {{Collado Ciprés}}, \citenamefont {Blomqvist},\
  and\ \citenamefont {Kaplan}}]{garcia_2019}%
  \BibitemOpen
  \bibfield  {author} {\bibinfo {author} {\bibfnamefont {J.}~\bibnamefont
  {García}}, \bibinfo {author} {\bibfnamefont {V.}~\bibnamefont {{Collado
  Ciprés}}}, \bibinfo {author} {\bibfnamefont {A.}~\bibnamefont {Blomqvist}},\
  and\ \bibinfo {author} {\bibfnamefont {B.}~\bibnamefont {Kaplan}},\
  }\bibfield  {title} {\bibinfo {title} {Cemented carbide microstructures: a
  review},\ }\href
  {https://doi.org/https://doi.org/10.1016/j.ijrmhm.2018.12.004} {\bibfield
  {journal} {\bibinfo  {journal} {International Journal of Refractory Metals
  and Hard Materials}\ }\textbf {\bibinfo {volume} {80}},\ \bibinfo {pages}
  {40} (\bibinfo {year} {2019})}\BibitemShut {NoStop}%
\bibitem [{\citenamefont {Eizadjou}\ \emph {et~al.}(2020)\citenamefont
  {Eizadjou}, \citenamefont {Chen}, \citenamefont {Czettl}, \citenamefont
  {Pachlhofer}, \citenamefont {Primig},\ and\ \citenamefont
  {Ringer}}]{eizadjou_2020}%
  \BibitemOpen
  \bibfield  {author} {\bibinfo {author} {\bibfnamefont {M.}~\bibnamefont
  {Eizadjou}}, \bibinfo {author} {\bibfnamefont {H.}~\bibnamefont {Chen}},
  \bibinfo {author} {\bibfnamefont {C.}~\bibnamefont {Czettl}}, \bibinfo
  {author} {\bibfnamefont {J.}~\bibnamefont {Pachlhofer}}, \bibinfo {author}
  {\bibfnamefont {S.}~\bibnamefont {Primig}},\ and\ \bibinfo {author}
  {\bibfnamefont {S.~P.}\ \bibnamefont {Ringer}},\ }\bibfield  {title}
  {\bibinfo {title} {An observation of the binder microstructure in
  {WC}-({Co}+{Ru}) cemented carbides using transmission {Kikuchi}
  diffraction},\ }\href {https://doi.org/10.1016/j.scriptamat.2020.03.010}
  {\bibfield  {journal} {\bibinfo  {journal} {Scripta Materialia}\ }\textbf
  {\bibinfo {volume} {183}},\ \bibinfo {pages} {55} (\bibinfo {year}
  {2020})}\BibitemShut {NoStop}%
\bibitem [{\citenamefont {Mayergoyz}(1986)}]{mayergoyz_1986}%
  \BibitemOpen
  \bibfield  {author} {\bibinfo {author} {\bibfnamefont {I.}~\bibnamefont
  {Mayergoyz}},\ }\bibfield  {title} {\bibinfo {title} {Mathematical models of
  hysteresis},\ }\href {https://doi.org/10.1109/TMAG.1986.1064347} {\bibfield
  {journal} {\bibinfo  {journal} {IEEE Transactions on Magnetics}\ }\textbf
  {\bibinfo {volume} {22}},\ \bibinfo {pages} {603} (\bibinfo {year} {1986})},\
  \bibinfo {note} {conference Name: IEEE Transactions on Magnetics}\BibitemShut
  {NoStop}%
\bibitem [{\citenamefont {Preisach}(1935)}]{preisach_1935}%
  \BibitemOpen
  \bibfield  {author} {\bibinfo {author} {\bibfnamefont {F.}~\bibnamefont
  {Preisach}},\ }\bibfield  {title} {\bibinfo {title} {\"uber die magnetische
  nachwirkung},\ }\href {https://doi.org/10.1007/BF01349418} {\bibfield
  {journal} {\bibinfo  {journal} {Z. Physik}\ }\textbf {\bibinfo {volume}
  {94}},\ \bibinfo {pages} {277} (\bibinfo {year} {1935})}\BibitemShut
  {NoStop}%
\bibitem [{\citenamefont {Torre}\ and\ \citenamefont
  {Bennett}(2005)}]{dellatorre_2005}%
  \BibitemOpen
  \bibfield  {author} {\bibinfo {author} {\bibfnamefont {E.~D.}\ \bibnamefont
  {Torre}}\ and\ \bibinfo {author} {\bibfnamefont {L.~H.}\ \bibnamefont
  {Bennett}},\ }\bibfield  {title} {\bibinfo {title} {Analysis and simulations
  of magnetic materials},\ }\href {https://doi.org/10.3934/proc.2005.2005.854}
  {\bibfield  {journal} {\bibinfo  {journal} {Conference Publications}\
  }\textbf {\bibinfo {volume} {2005}},\ \bibinfo {pages} {854} (\bibinfo {year}
  {2005})},\ \bibinfo {note} {company: Conference Publications Distributor:
  Conference Publications Institution: Conference Publications Label:
  Conference Publications Publisher: American Institute of Mathematical
  Sciences}\BibitemShut {NoStop}%
\bibitem [{\citenamefont {Della~Torre}(1966)}]{dellatorre_1966}%
  \BibitemOpen
  \bibfield  {author} {\bibinfo {author} {\bibfnamefont {E.}~\bibnamefont
  {Della~Torre}},\ }\bibfield  {title} {\bibinfo {title} {Effect of interaction
  on the magnetization of single-domain particles},\ }\href
  {https://doi.org/10.1109/TAU.1966.1161852} {\bibfield  {journal} {\bibinfo
  {journal} {IEEE Transactions on Audio and Electroacoustics}\ }\textbf
  {\bibinfo {volume} {14}},\ \bibinfo {pages} {86} (\bibinfo {year}
  {1966})}\BibitemShut {NoStop}%
\bibitem [{\citenamefont {Kádár}(1987)}]{kadar_1987}%
  \BibitemOpen
  \bibfield  {author} {\bibinfo {author} {\bibfnamefont {G.}~\bibnamefont
  {Kádár}},\ }\bibfield  {title} {\bibinfo {title} {On the preisach function
  of ferromagnetic hysteresis},\ }\href {https://doi.org/10.1063/1.338563}
  {\bibfield  {journal} {\bibinfo  {journal} {Journal of Applied Physics}\
  }\textbf {\bibinfo {volume} {61}},\ \bibinfo {pages} {4013} (\bibinfo {year}
  {1987})},\ \Eprint {https://arxiv.org/abs/https://doi.org/10.1063/1.338563}
  {https://doi.org/10.1063/1.338563} \BibitemShut {NoStop}%
\bibitem [{\citenamefont {Mayergoyz}\ and\ \citenamefont
  {Friedman}(1988)}]{mayergoyz_1988}%
  \BibitemOpen
  \bibfield  {author} {\bibinfo {author} {\bibfnamefont {I.}~\bibnamefont
  {Mayergoyz}}\ and\ \bibinfo {author} {\bibfnamefont {G.}~\bibnamefont
  {Friedman}},\ }\bibfield  {title} {\bibinfo {title} {Generalized preisach
  model of hysteresis},\ }\href {https://doi.org/10.1109/20.43892} {\bibfield
  {journal} {\bibinfo  {journal} {IEEE Transactions on Magnetics}\ }\textbf
  {\bibinfo {volume} {24}},\ \bibinfo {pages} {212} (\bibinfo {year}
  {1988})}\BibitemShut {NoStop}%
\bibitem [{\citenamefont {Egli}(2021)}]{egli_2021}%
  \BibitemOpen
  \bibfield  {author} {\bibinfo {author} {\bibfnamefont {R.}~\bibnamefont
  {Egli}},\ }\bibinfo {title} {Magnetic characterization of geologic materials
  with first-order reversal curves},\ in\ \href
  {https://doi.org/10.1007/978-3-030-70443-8_17} {\emph {\bibinfo {booktitle}
  {Magnetic Measurement Techniques for Materials Characterization}}},\ \bibinfo
  {editor} {edited by\ \bibinfo {editor} {\bibfnamefont {V.}~\bibnamefont
  {Franco}}\ and\ \bibinfo {editor} {\bibfnamefont {B.}~\bibnamefont
  {Dodrill}}}\ (\bibinfo  {publisher} {Springer International Publishing},\
  \bibinfo {address} {Cham},\ \bibinfo {year} {2021})\ pp.\ \bibinfo {pages}
  {455--604}\BibitemShut {NoStop}%
\bibitem [{\citenamefont {Pike}\ \emph {et~al.}(1999)\citenamefont {Pike},
  \citenamefont {Roberts},\ and\ \citenamefont {Verosub}}]{pike_1999}%
  \BibitemOpen
  \bibfield  {author} {\bibinfo {author} {\bibfnamefont {C.~R.}\ \bibnamefont
  {Pike}}, \bibinfo {author} {\bibfnamefont {A.~P.}\ \bibnamefont {Roberts}},\
  and\ \bibinfo {author} {\bibfnamefont {K.~L.}\ \bibnamefont {Verosub}},\
  }\bibfield  {title} {\bibinfo {title} {Characterizing interactions in fine
  magnetic particle systems using first order reversal curves},\ }\href
  {https://doi.org/10.1063/1.370176} {\bibfield  {journal} {\bibinfo  {journal}
  {Journal of Applied Physics}\ }\textbf {\bibinfo {volume} {85}},\ \bibinfo
  {pages} {6660} (\bibinfo {year} {1999})},\ \bibinfo {note} {publisher:
  American Institute of Physics}\BibitemShut {NoStop}%
\bibitem [{\citenamefont {Pike}\ \emph {et~al.}(2005)\citenamefont {Pike},
  \citenamefont {Ross}, \citenamefont {Scalettar},\ and\ \citenamefont
  {Zimanyi}}]{pike_2005}%
  \BibitemOpen
  \bibfield  {author} {\bibinfo {author} {\bibfnamefont {C.~R.}\ \bibnamefont
  {Pike}}, \bibinfo {author} {\bibfnamefont {C.~A.}\ \bibnamefont {Ross}},
  \bibinfo {author} {\bibfnamefont {R.~T.}\ \bibnamefont {Scalettar}},\ and\
  \bibinfo {author} {\bibfnamefont {G.}~\bibnamefont {Zimanyi}},\ }\bibfield
  {title} {\bibinfo {title} {First-order reversal curve diagram analysis of a
  perpendicular nickel nanopillar array},\ }\href
  {https://doi.org/10.1103/PhysRevB.71.134407} {\bibfield  {journal} {\bibinfo
  {journal} {Phys. Rev. B}\ }\textbf {\bibinfo {volume} {71}},\ \bibinfo
  {pages} {134407} (\bibinfo {year} {2005})},\ \bibinfo {note} {publisher:
  American Physical Society}\BibitemShut {NoStop}%
\bibitem [{\citenamefont {Ross}\ \emph {et~al.}(2002)\citenamefont {Ross},
  \citenamefont {Hwang}, \citenamefont {Shima}, \citenamefont {Cheng},
  \citenamefont {Farhoud}, \citenamefont {Savas}, \citenamefont {Smith},
  \citenamefont {Schwarzacher}, \citenamefont {Ross}, \citenamefont {Redjdal},\
  and\ \citenamefont {Humphrey}}]{ross_2002}%
  \BibitemOpen
  \bibfield  {author} {\bibinfo {author} {\bibfnamefont {C.~A.}\ \bibnamefont
  {Ross}}, \bibinfo {author} {\bibfnamefont {M.}~\bibnamefont {Hwang}},
  \bibinfo {author} {\bibfnamefont {M.}~\bibnamefont {Shima}}, \bibinfo
  {author} {\bibfnamefont {J.~Y.}\ \bibnamefont {Cheng}}, \bibinfo {author}
  {\bibfnamefont {M.}~\bibnamefont {Farhoud}}, \bibinfo {author} {\bibfnamefont
  {T.~A.}\ \bibnamefont {Savas}}, \bibinfo {author} {\bibfnamefont {H.~I.}\
  \bibnamefont {Smith}}, \bibinfo {author} {\bibfnamefont {W.}~\bibnamefont
  {Schwarzacher}}, \bibinfo {author} {\bibfnamefont {F.~M.}\ \bibnamefont
  {Ross}}, \bibinfo {author} {\bibfnamefont {M.}~\bibnamefont {Redjdal}},\ and\
  \bibinfo {author} {\bibfnamefont {F.~B.}\ \bibnamefont {Humphrey}},\
  }\bibfield  {title} {\bibinfo {title} {Micromagnetic behavior of
  electrodeposited cylinder arrays},\ }\href
  {https://doi.org/10.1103/PhysRevB.65.144417} {\bibfield  {journal} {\bibinfo
  {journal} {Phys. Rev. B}\ }\textbf {\bibinfo {volume} {65}},\ \bibinfo
  {pages} {144417} (\bibinfo {year} {2002})},\ \bibinfo {note} {publisher:
  American Physical Society}\BibitemShut {NoStop}%
\bibitem [{\citenamefont {Valdez-Grijalva}\ \emph {et~al.}(2020)\citenamefont
  {Valdez-Grijalva}, \citenamefont {Nagy}, \citenamefont {Muxworthy},
  \citenamefont {Williams}, \citenamefont {Roberts},\ and\ \citenamefont
  {Heslop}}]{valdez_2020}%
  \BibitemOpen
  \bibfield  {author} {\bibinfo {author} {\bibfnamefont {M.~A.}\ \bibnamefont
  {Valdez-Grijalva}}, \bibinfo {author} {\bibfnamefont {L.}~\bibnamefont
  {Nagy}}, \bibinfo {author} {\bibfnamefont {A.~R.}\ \bibnamefont {Muxworthy}},
  \bibinfo {author} {\bibfnamefont {W.}~\bibnamefont {Williams}}, \bibinfo
  {author} {\bibfnamefont {A.~P.}\ \bibnamefont {Roberts}},\ and\ \bibinfo
  {author} {\bibfnamefont {D.}~\bibnamefont {Heslop}},\ }\bibfield  {title}
  {\bibinfo {title} {{Micromagnetic simulations of first-order reversal curve
  (FORC) diagrams of framboidal greigite}},\ }\href
  {https://doi.org/10.1093/gji/ggaa241} {\bibfield  {journal} {\bibinfo
  {journal} {Geophysical Journal International}\ }\textbf {\bibinfo {volume}
  {222}},\ \bibinfo {pages} {1126} (\bibinfo {year} {2020})},\ \Eprint
  {https://arxiv.org/abs/https://academic.oup.com/gji/article-pdf/222/2/1126/33331481/ggaa241.pdf}
  {https://academic.oup.com/gji/article-pdf/222/2/1126/33331481/ggaa241.pdf}
  \BibitemShut {NoStop}%
\bibitem [{\citenamefont {Nagy}\ \emph {et~al.}(2019)\citenamefont {Nagy},
  \citenamefont {Williams}, \citenamefont {Tauxe},\ and\ \citenamefont
  {Muxworthy}}]{nagy_2019}%
  \BibitemOpen
  \bibfield  {author} {\bibinfo {author} {\bibfnamefont {L.}~\bibnamefont
  {Nagy}}, \bibinfo {author} {\bibfnamefont {W.}~\bibnamefont {Williams}},
  \bibinfo {author} {\bibfnamefont {L.}~\bibnamefont {Tauxe}},\ and\ \bibinfo
  {author} {\bibfnamefont {A.~R.}\ \bibnamefont {Muxworthy}},\ }\bibfield
  {title} {\bibinfo {title} {From nano to micro: Evolution of magnetic domain
  structures in multidomain magnetite},\ }\href
  {https://doi.org/https://doi.org/10.1029/2019GC008319} {\bibfield  {journal}
  {\bibinfo  {journal} {Geochemistry, Geophysics, Geosystems}\ }\textbf
  {\bibinfo {volume} {20}},\ \bibinfo {pages} {2907} (\bibinfo {year}
  {2019})},\ \Eprint
  {https://arxiv.org/abs/https://agupubs.onlinelibrary.wiley.com/doi/pdf/10.1029/2019GC008319}
  {https://agupubs.onlinelibrary.wiley.com/doi/pdf/10.1029/2019GC008319}
  \BibitemShut {NoStop}%
\bibitem [{\citenamefont {Fischbacher}\ \emph {et~al.}(2017)\citenamefont
  {Fischbacher}, \citenamefont {Kovacs}, \citenamefont {Oezelt}, \citenamefont
  {Schrefl}, \citenamefont {Exl}, \citenamefont {Fidler}, \citenamefont
  {Suess}, \citenamefont {Sakuma}, \citenamefont {Yano}, \citenamefont {Kato},
  \citenamefont {Shoji},\ and\ \citenamefont {Manabe}}]{fischbacher_2017}%
  \BibitemOpen
  \bibfield  {author} {\bibinfo {author} {\bibfnamefont {J.}~\bibnamefont
  {Fischbacher}}, \bibinfo {author} {\bibfnamefont {A.}~\bibnamefont {Kovacs}},
  \bibinfo {author} {\bibfnamefont {H.}~\bibnamefont {Oezelt}}, \bibinfo
  {author} {\bibfnamefont {T.}~\bibnamefont {Schrefl}}, \bibinfo {author}
  {\bibfnamefont {L.}~\bibnamefont {Exl}}, \bibinfo {author} {\bibfnamefont
  {J.}~\bibnamefont {Fidler}}, \bibinfo {author} {\bibfnamefont
  {D.}~\bibnamefont {Suess}}, \bibinfo {author} {\bibfnamefont
  {N.}~\bibnamefont {Sakuma}}, \bibinfo {author} {\bibfnamefont
  {M.}~\bibnamefont {Yano}}, \bibinfo {author} {\bibfnamefont {A.}~\bibnamefont
  {Kato}}, \bibinfo {author} {\bibfnamefont {T.}~\bibnamefont {Shoji}},\ and\
  \bibinfo {author} {\bibfnamefont {A.}~\bibnamefont {Manabe}},\ }\bibfield
  {title} {\bibinfo {title} {Nonlinear conjugate gradient methods in
  micromagnetics},\ }\href {https://doi.org/10.1063/1.4981902} {\bibfield
  {journal} {\bibinfo  {journal} {AIP Advances}\ }\textbf {\bibinfo {volume}
  {7}},\ \bibinfo {pages} {045310} (\bibinfo {year} {2017})},\ \bibinfo {note}
  {publisher: American Institute of Physics}\BibitemShut {NoStop}%
\bibitem [{\citenamefont {Schrefl}\ \emph {et~al.}(2007)\citenamefont
  {Schrefl}, \citenamefont {Hrkac}, \citenamefont {Bance}, \citenamefont
  {Suess}, \citenamefont {Ertl},\ and\ \citenamefont {Fidler}}]{schrefl_2007}%
  \BibitemOpen
  \bibfield  {author} {\bibinfo {author} {\bibfnamefont {T.}~\bibnamefont
  {Schrefl}}, \bibinfo {author} {\bibfnamefont {G.}~\bibnamefont {Hrkac}},
  \bibinfo {author} {\bibfnamefont {S.}~\bibnamefont {Bance}}, \bibinfo
  {author} {\bibfnamefont {D.}~\bibnamefont {Suess}}, \bibinfo {author}
  {\bibfnamefont {O.}~\bibnamefont {Ertl}},\ and\ \bibinfo {author}
  {\bibfnamefont {J.}~\bibnamefont {Fidler}},\ }\bibinfo {title} {Numerical
  methods in micromagnetics (finite element method)},\ in\ \href
  {https://doi.org/10.1002/9780470022184} {\emph {\bibinfo {booktitle}
  {Handbook of Magnetism and Advanced Magnetic Materials}}},\ Vol.\ \bibinfo
  {volume} {Micromagnetism}\ (\bibinfo  {publisher} {John Wiley \& Sons,
  Ltd.},\ \bibinfo {year} {2007})\BibitemShut {NoStop}%
\bibitem [{\citenamefont {Exl}\ \emph {et~al.}(2018)\citenamefont {Exl},
  \citenamefont {Fischbacher}, \citenamefont {Kovacs}, \citenamefont {Özelt},
  \citenamefont {Gusenbauer}, \citenamefont {Yokota}, \citenamefont {Shoji},
  \citenamefont {Hrkac},\ and\ \citenamefont {Schrefl}}]{exl_2018}%
  \BibitemOpen
  \bibfield  {author} {\bibinfo {author} {\bibfnamefont {L.}~\bibnamefont
  {Exl}}, \bibinfo {author} {\bibfnamefont {J.}~\bibnamefont {Fischbacher}},
  \bibinfo {author} {\bibfnamefont {A.}~\bibnamefont {Kovacs}}, \bibinfo
  {author} {\bibfnamefont {H.}~\bibnamefont {Özelt}}, \bibinfo {author}
  {\bibfnamefont {M.}~\bibnamefont {Gusenbauer}}, \bibinfo {author}
  {\bibfnamefont {K.}~\bibnamefont {Yokota}}, \bibinfo {author} {\bibfnamefont
  {T.}~\bibnamefont {Shoji}}, \bibinfo {author} {\bibfnamefont
  {G.}~\bibnamefont {Hrkac}},\ and\ \bibinfo {author} {\bibfnamefont
  {T.}~\bibnamefont {Schrefl}},\ }\bibfield  {title} {\bibinfo {title}
  {Magnetic microstructure machine learning analysis},\ }\bibfield  {journal}
  {\bibinfo  {journal} {J. Phys. Mater.}\ }\href
  {https://doi.org/10.1088/2515-7639/aaf26d} {10.1088/2515-7639/aaf26d}
  (\bibinfo {year} {2018}),\ \bibinfo {note} {publisher: IOP
  Publishing}\BibitemShut {NoStop}%
\bibitem [{\citenamefont {Exl}\ \emph {et~al.}(2020)\citenamefont {Exl},
  \citenamefont {Suess},\ and\ \citenamefont {Schrefl}}]{Exl_2020}%
  \BibitemOpen
  \bibfield  {author} {\bibinfo {author} {\bibfnamefont {L.}~\bibnamefont
  {Exl}}, \bibinfo {author} {\bibfnamefont {D.}~\bibnamefont {Suess}},\ and\
  \bibinfo {author} {\bibfnamefont {T.}~\bibnamefont {Schrefl}},\ }\bibinfo
  {title} {Micromagnetism},\ in\ \href
  {https://doi.org/10.1007/978-3-030-63101-7_7-1} {\emph {\bibinfo {booktitle}
  {Handbook of Magnetism and Magnetic Materials}}},\ \bibinfo {editor} {edited
  by\ \bibinfo {editor} {\bibfnamefont {M.}~\bibnamefont {Coey}}\ and\ \bibinfo
  {editor} {\bibfnamefont {S.}~\bibnamefont {Parkin}}}\ (\bibinfo  {publisher}
  {Springer International Publishing},\ \bibinfo {address} {Cham},\ \bibinfo
  {year} {2020})\ pp.\ \bibinfo {pages} {1--44}\BibitemShut {NoStop}%
\bibitem [{\citenamefont {Pike}\ and\ \citenamefont
  {Fernandez}(1999)}]{pike_fernandez_1999}%
  \BibitemOpen
  \bibfield  {author} {\bibinfo {author} {\bibfnamefont {C.}~\bibnamefont
  {Pike}}\ and\ \bibinfo {author} {\bibfnamefont {A.}~\bibnamefont
  {Fernandez}},\ }\bibfield  {title} {\bibinfo {title} {An investigation of
  magnetic reversal in submicron-scale {Co} dots using first order reversal
  curve diagrams},\ }\href {https://doi.org/10.1063/1.370177} {\bibfield
  {journal} {\bibinfo  {journal} {Journal of Applied Physics}\ }\textbf
  {\bibinfo {volume} {85}},\ \bibinfo {pages} {6668} (\bibinfo {year}
  {1999})}\BibitemShut {NoStop}%
\bibitem [{\citenamefont {Rave}\ \emph {et~al.}(1998)\citenamefont {Rave},
  \citenamefont {Fabian},\ and\ \citenamefont {Hubert}}]{rave_1998}%
  \BibitemOpen
  \bibfield  {author} {\bibinfo {author} {\bibfnamefont {W.}~\bibnamefont
  {Rave}}, \bibinfo {author} {\bibfnamefont {K.}~\bibnamefont {Fabian}},\ and\
  \bibinfo {author} {\bibfnamefont {A.}~\bibnamefont {Hubert}},\ }\bibfield
  {title} {\bibinfo {title} {Magnetic states of small cubic particles with
  uniaxial anisotropy},\ }\href
  {https://doi.org/https://doi.org/10.1016/S0304-8853(98)00328-X} {\bibfield
  {journal} {\bibinfo  {journal} {Journal of Magnetism and Magnetic Materials}\
  }\textbf {\bibinfo {volume} {190}},\ \bibinfo {pages} {332} (\bibinfo {year}
  {1998})}\BibitemShut {NoStop}%
\bibitem [{\citenamefont {Groß}\ \emph {et~al.}(2019)\citenamefont {Groß},
  \citenamefont {Ilse}, \citenamefont {Schütz}, \citenamefont {Gräfe},\ and\
  \citenamefont {Goering}}]{gross_2019}%
  \BibitemOpen
  \bibfield  {author} {\bibinfo {author} {\bibfnamefont {F.}~\bibnamefont
  {Groß}}, \bibinfo {author} {\bibfnamefont {S.~E.}\ \bibnamefont {Ilse}},
  \bibinfo {author} {\bibfnamefont {G.}~\bibnamefont {Schütz}}, \bibinfo
  {author} {\bibfnamefont {J.}~\bibnamefont {Gräfe}},\ and\ \bibinfo {author}
  {\bibfnamefont {E.}~\bibnamefont {Goering}},\ }\bibfield  {title} {\bibinfo
  {title} {Interpreting first-order reversal curves beyond the {Preisach}
  model: {An} experimental permalloy microarray investigation},\ }\href
  {https://doi.org/10.1103/PhysRevB.99.064401} {\bibfield  {journal} {\bibinfo
  {journal} {Phys. Rev. B}\ }\textbf {\bibinfo {volume} {99}},\ \bibinfo
  {pages} {064401} (\bibinfo {year} {2019})},\ \bibinfo {note} {publisher:
  American Physical Society}\BibitemShut {NoStop}%
\bibitem [{\citenamefont {Muxworthy}\ \emph {et~al.}(2004)\citenamefont
  {Muxworthy}, \citenamefont {Heslop},\ and\ \citenamefont
  {Williams}}]{muxworthy_2004}%
  \BibitemOpen
  \bibfield  {author} {\bibinfo {author} {\bibfnamefont {A.}~\bibnamefont
  {Muxworthy}}, \bibinfo {author} {\bibfnamefont {D.}~\bibnamefont {Heslop}},\
  and\ \bibinfo {author} {\bibfnamefont {W.}~\bibnamefont {Williams}},\
  }\bibfield  {title} {\bibinfo {title} {{Influence of magnetostatic
  interactions on first-order-reversal-curve (FORC) diagrams: a micromagnetic
  approach}},\ }\href {https://doi.org/10.1111/j.1365-246X.2004.02358.x}
  {\bibfield  {journal} {\bibinfo  {journal} {Geophysical Journal
  International}\ }\textbf {\bibinfo {volume} {158}},\ \bibinfo {pages} {888}
  (\bibinfo {year} {2004})},\ \Eprint
  {https://arxiv.org/abs/https://academic.oup.com/gji/article-pdf/158/3/888/5987716/158-3-888.pdf}
  {https://academic.oup.com/gji/article-pdf/158/3/888/5987716/158-3-888.pdf}
  \BibitemShut {NoStop}%
\bibitem [{\citenamefont {Stancu}\ \emph {et~al.}(2003)\citenamefont {Stancu},
  \citenamefont {Pike}, \citenamefont {Stoleriu}, \citenamefont {Postolache},\
  and\ \citenamefont {Cimpoesu}}]{stancu_2003}%
  \BibitemOpen
  \bibfield  {author} {\bibinfo {author} {\bibfnamefont {A.}~\bibnamefont
  {Stancu}}, \bibinfo {author} {\bibfnamefont {C.}~\bibnamefont {Pike}},
  \bibinfo {author} {\bibfnamefont {L.}~\bibnamefont {Stoleriu}}, \bibinfo
  {author} {\bibfnamefont {P.}~\bibnamefont {Postolache}},\ and\ \bibinfo
  {author} {\bibfnamefont {D.}~\bibnamefont {Cimpoesu}},\ }\bibfield  {title}
  {\bibinfo {title} {Micromagnetic and preisach analysis of the first order
  reversal curves (forc) diagram},\ }\href {https://doi.org/10.1063/1.1557656}
  {\bibfield  {journal} {\bibinfo  {journal} {Journal of Applied Physics}\
  }\textbf {\bibinfo {volume} {93}},\ \bibinfo {pages} {6620} (\bibinfo {year}
  {2003})},\ \Eprint {https://arxiv.org/abs/https://doi.org/10.1063/1.1557656}
  {https://doi.org/10.1063/1.1557656} \BibitemShut {NoStop}%
\bibitem [{\citenamefont {Pike}\ \emph {et~al.}(2001)\citenamefont {Pike},
  \citenamefont {Roberts}, \citenamefont {Dekkers},\ and\ \citenamefont
  {Verosub}}]{pike_2001}%
  \BibitemOpen
  \bibfield  {author} {\bibinfo {author} {\bibfnamefont {C.~R.}\ \bibnamefont
  {Pike}}, \bibinfo {author} {\bibfnamefont {A.~P.}\ \bibnamefont {Roberts}},
  \bibinfo {author} {\bibfnamefont {M.~J.}\ \bibnamefont {Dekkers}},\ and\
  \bibinfo {author} {\bibfnamefont {K.~L.}\ \bibnamefont {Verosub}},\
  }\bibfield  {title} {\bibinfo {title} {An investigation of multi-domain
  hysteresis mechanisms using forc diagrams},\ }\href
  {https://doi.org/https://doi.org/10.1016/S0031-9201(01)00241-2} {\bibfield
  {journal} {\bibinfo  {journal} {Physics of the Earth and Planetary
  Interiors}\ }\textbf {\bibinfo {volume} {126}},\ \bibinfo {pages} {11}
  (\bibinfo {year} {2001})},\ \bibinfo {note} {rock Magnetism Enters the New
  Millenium. A Celebration of Fifty Years of Neel's Theories}\BibitemShut
  {NoStop}%
\bibitem [{\citenamefont {Dunlop}\ \emph {et~al.}(1990)\citenamefont {Dunlop},
  \citenamefont {Westcott-Lewis},\ and\ \citenamefont {Bailey}}]{dunlop_1990}%
  \BibitemOpen
  \bibfield  {author} {\bibinfo {author} {\bibfnamefont {D.~J.}\ \bibnamefont
  {Dunlop}}, \bibinfo {author} {\bibfnamefont {M.~F.}\ \bibnamefont
  {Westcott-Lewis}},\ and\ \bibinfo {author} {\bibfnamefont {M.~E.}\
  \bibnamefont {Bailey}},\ }\bibfield  {title} {\bibinfo {title} {Preisach
  diagrams and anhysteresis: do they measure interactions?},\ }\href
  {https://doi.org/https://doi.org/10.1016/0031-9201(90)90076-A} {\bibfield
  {journal} {\bibinfo  {journal} {Physics of the Earth and Planetary
  Interiors}\ }\textbf {\bibinfo {volume} {65}},\ \bibinfo {pages} {62}
  (\bibinfo {year} {1990})}\BibitemShut {NoStop}%
\bibitem [{\citenamefont {Lascu}\ \emph {et~al.}(2018)\citenamefont {Lascu},
  \citenamefont {Einsle}, \citenamefont {Ball},\ and\ \citenamefont
  {Harrison}}]{lascu_2018}%
  \BibitemOpen
  \bibfield  {author} {\bibinfo {author} {\bibfnamefont {I.}~\bibnamefont
  {Lascu}}, \bibinfo {author} {\bibfnamefont {J.~F.}\ \bibnamefont {Einsle}},
  \bibinfo {author} {\bibfnamefont {M.~R.}\ \bibnamefont {Ball}},\ and\
  \bibinfo {author} {\bibfnamefont {R.~J.}\ \bibnamefont {Harrison}},\
  }\bibfield  {title} {\bibinfo {title} {The vortex state in geologic
  materials: A micromagnetic perspective},\ }\href
  {https://doi.org/https://doi.org/10.1029/2018JB015909} {\bibfield  {journal}
  {\bibinfo  {journal} {Journal of Geophysical Research: Solid Earth}\ }\textbf
  {\bibinfo {volume} {123}},\ \bibinfo {pages} {7285} (\bibinfo {year}
  {2018})}\BibitemShut {NoStop}%
\bibitem [{\citenamefont {Roberts}\ \emph {et~al.}(2017)\citenamefont
  {Roberts}, \citenamefont {Almeida}, \citenamefont {Church}, \citenamefont
  {Harrison}, \citenamefont {Heslop}, \citenamefont {Li}, \citenamefont {Li},
  \citenamefont {Muxworthy}, \citenamefont {Williams},\ and\ \citenamefont
  {Zhao}}]{roberts_2017}%
  \BibitemOpen
  \bibfield  {author} {\bibinfo {author} {\bibfnamefont {A.~P.}\ \bibnamefont
  {Roberts}}, \bibinfo {author} {\bibfnamefont {T.~P.}\ \bibnamefont
  {Almeida}}, \bibinfo {author} {\bibfnamefont {N.~S.}\ \bibnamefont {Church}},
  \bibinfo {author} {\bibfnamefont {R.~J.}\ \bibnamefont {Harrison}}, \bibinfo
  {author} {\bibfnamefont {D.}~\bibnamefont {Heslop}}, \bibinfo {author}
  {\bibfnamefont {Y.}~\bibnamefont {Li}}, \bibinfo {author} {\bibfnamefont
  {J.}~\bibnamefont {Li}}, \bibinfo {author} {\bibfnamefont {A.~R.}\
  \bibnamefont {Muxworthy}}, \bibinfo {author} {\bibfnamefont {W.}~\bibnamefont
  {Williams}},\ and\ \bibinfo {author} {\bibfnamefont {X.}~\bibnamefont
  {Zhao}},\ }\bibfield  {title} {\bibinfo {title} {Resolving the origin of
  pseudo-single domain magnetic behavior},\ }\href
  {https://doi.org/https://doi.org/10.1002/2017JB014860} {\bibfield  {journal}
  {\bibinfo  {journal} {Journal of Geophysical Research: Solid Earth}\ }\textbf
  {\bibinfo {volume} {122}},\ \bibinfo {pages} {9534} (\bibinfo {year}
  {2017})}\BibitemShut {NoStop}%
\bibitem [{\citenamefont {Carvallo}\ \emph {et~al.}(2003)\citenamefont
  {Carvallo}, \citenamefont {Muxworthy}, \citenamefont {Dunlop},\ and\
  \citenamefont {Williams}}]{carvallo_2003}%
  \BibitemOpen
  \bibfield  {author} {\bibinfo {author} {\bibfnamefont {C.}~\bibnamefont
  {Carvallo}}, \bibinfo {author} {\bibfnamefont {A.~R.}\ \bibnamefont
  {Muxworthy}}, \bibinfo {author} {\bibfnamefont {D.~J.}\ \bibnamefont
  {Dunlop}},\ and\ \bibinfo {author} {\bibfnamefont {W.}~\bibnamefont
  {Williams}},\ }\bibfield  {title} {\bibinfo {title} {Micromagnetic modeling
  of first-order reversal curve (forc) diagrams for single-domain and
  pseudo-single-domain magnetite},\ }\href
  {https://doi.org/https://doi.org/10.1016/S0012-821X(03)00320-0} {\bibfield
  {journal} {\bibinfo  {journal} {Earth and Planetary Science Letters}\
  }\textbf {\bibinfo {volume} {213}},\ \bibinfo {pages} {375} (\bibinfo {year}
  {2003})}\BibitemShut {NoStop}%
\bibitem [{\citenamefont {Harrison}\ and\ \citenamefont
  {Feinberg}(2008)}]{harrison_forcinel_2008}%
  \BibitemOpen
  \bibfield  {author} {\bibinfo {author} {\bibfnamefont {R.~J.}\ \bibnamefont
  {Harrison}}\ and\ \bibinfo {author} {\bibfnamefont {J.~M.}\ \bibnamefont
  {Feinberg}},\ }\bibfield  {title} {\bibinfo {title} {{FORCinel}: {An}
  improved algorithm for calculating first-order reversal curve distributions
  using locally weighted regression smoothing},\ }\bibfield  {journal}
  {\bibinfo  {journal} {Geochemistry, Geophysics, Geosystems}\ }\textbf
  {\bibinfo {volume} {9}},\ \href {https://doi.org/10.1029/2008GC001987}
  {10.1029/2008GC001987} (\bibinfo {year} {2008}),\ \bibinfo {note} {\_eprint:
  https://onlinelibrary.wiley.com/doi/pdf/10.1029/2008GC001987}\BibitemShut
  {NoStop}%
\bibitem [{\citenamefont {Egli}(2013)}]{egli_variforc_2013}%
  \BibitemOpen
  \bibfield  {author} {\bibinfo {author} {\bibfnamefont {R.}~\bibnamefont
  {Egli}},\ }\bibfield  {title} {\bibinfo {title} {{VARIFORC}: {An} optimized
  protocol for calculating non-regular first-order reversal curve ({FORC})
  diagrams},\ }\href {https://doi.org/10.1016/j.gloplacha.2013.08.003}
  {\bibfield  {journal} {\bibinfo  {journal} {Global and Planetary Change}\
  }\bibinfo {series} {Magnetic iron minerals in sediments and their relation to
  geologic processes, climate, and the geomagnetic field},\ \textbf {\bibinfo
  {volume} {110}},\ \bibinfo {pages} {302} (\bibinfo {year}
  {2013})}\BibitemShut {NoStop}%
\bibitem [{\citenamefont {Zhao}\ \emph {et~al.}(2017)\citenamefont {Zhao},
  \citenamefont {Roberts}, \citenamefont {Heslop}, \citenamefont {Paterson},
  \citenamefont {Li},\ and\ \citenamefont {Li}}]{zhao_2017}%
  \BibitemOpen
  \bibfield  {author} {\bibinfo {author} {\bibfnamefont {X.}~\bibnamefont
  {Zhao}}, \bibinfo {author} {\bibfnamefont {A.~P.}\ \bibnamefont {Roberts}},
  \bibinfo {author} {\bibfnamefont {D.}~\bibnamefont {Heslop}}, \bibinfo
  {author} {\bibfnamefont {G.~A.}\ \bibnamefont {Paterson}}, \bibinfo {author}
  {\bibfnamefont {Y.}~\bibnamefont {Li}},\ and\ \bibinfo {author}
  {\bibfnamefont {J.}~\bibnamefont {Li}},\ }\bibfield  {title} {\bibinfo
  {title} {Magnetic domain state diagnosis using hysteresis reversal curves},\
  }\href {https://doi.org/10.1002/2016JB013683} {\bibfield  {journal} {\bibinfo
   {journal} {Journal of Geophysical Research: Solid Earth}\ }\textbf {\bibinfo
  {volume} {122}},\ \bibinfo {pages} {4767} (\bibinfo {year} {2017})},\
  \bibinfo {note} {\_eprint:
  https://onlinelibrary.wiley.com/doi/pdf/10.1002/2016JB013683}\BibitemShut
  {NoStop}%
\bibitem [{\citenamefont {Visscher}(2019)}]{visscher_2019}%
  \BibitemOpen
  \bibfield  {author} {\bibinfo {author} {\bibfnamefont {P.~B.}\ \bibnamefont
  {Visscher}},\ }\bibfield  {title} {\bibinfo {title} {Avoiding the
  zero-coercivity anomaly in first order reversal curves: {FORC}+},\ }\href
  {https://doi.org/10.1063/1.5080101} {\bibfield  {journal} {\bibinfo
  {journal} {AIP Advances}\ }\textbf {\bibinfo {volume} {9}},\ \bibinfo {pages}
  {035117} (\bibinfo {year} {2019})},\ \bibinfo {note} {publisher: American
  Institute of Physics}\BibitemShut {NoStop}%
\bibitem [{\citenamefont {Berndt}\ and\ \citenamefont
  {Chang}(2019)}]{berndt_2019}%
  \BibitemOpen
  \bibfield  {author} {\bibinfo {author} {\bibfnamefont {T.~A.}\ \bibnamefont
  {Berndt}}\ and\ \bibinfo {author} {\bibfnamefont {L.}~\bibnamefont {Chang}},\
  }\bibfield  {title} {\bibinfo {title} {Waiting for {Forcot}: {Accelerating}
  {FORC} {Processing} 100× {Using} a {Fast}-{Fourier}-{Transform}
  {Algorithm}},\ }\href {https://doi.org/10.1029/2019GC008380} {\bibfield
  {journal} {\bibinfo  {journal} {Geochemistry, Geophysics, Geosystems}\
  }\textbf {\bibinfo {volume} {20}},\ \bibinfo {pages} {6223} (\bibinfo {year}
  {2019})},\ \bibinfo {note} {\_eprint:
  https://onlinelibrary.wiley.com/doi/pdf/10.1029/2019GC008380}\BibitemShut
  {NoStop}%
\bibitem [{\citenamefont {Cimpoesu}\ \emph {et~al.}(2019)\citenamefont
  {Cimpoesu}, \citenamefont {Dumitru},\ and\ \citenamefont
  {Stancu}}]{cimpoesu_2019}%
  \BibitemOpen
  \bibfield  {author} {\bibinfo {author} {\bibfnamefont {D.}~\bibnamefont
  {Cimpoesu}}, \bibinfo {author} {\bibfnamefont {I.}~\bibnamefont {Dumitru}},\
  and\ \bibinfo {author} {\bibfnamefont {A.}~\bibnamefont {Stancu}},\
  }\bibfield  {title} {\bibinfo {title} {doforc tool for calculating
  first-order reversal curve diagrams of noisy scattered data},\ }\href
  {https://doi.org/10.1063/1.5066445} {\bibfield  {journal} {\bibinfo
  {journal} {Journal of Applied Physics}\ }\textbf {\bibinfo {volume} {125}},\
  \bibinfo {pages} {023906} (\bibinfo {year} {2019})}\BibitemShut {NoStop}%
\end{thebibliography}%

\end{document}